\documentclass[conference,10pt]{IEEEtran}

\usepackage{framed}
\usepackage{tcolorbox}
\usepackage{algorithm}
\usepackage[noend]{algpseudocode}
\usepackage{bm}
\usepackage{stackrel}
\usepackage{subfigure}
\usepackage{epsf}
\usepackage{amsmath,amssymb}
\usepackage{graphicx}
\usepackage{color}
\usepackage{cite}
\usepackage{multirow,tabularx}
\usepackage{ifthen}
\usepackage{epstopdf}
\usepackage{mathtools}
\input{epstopdf.sty}
\usepackage{times}

\input{epsf.sty}
\makeatletter
\def\BState{\State\hskip-\ALG@thistlm}
\makeatother

\newcommand{\hide}[1]{\ifthenelse{\boolean{false}}{#1}{}}

\newtheorem{theorem}{{\bf Theorem}}

\newtheorem{lemma}{{\bf Lemma}}

\newenvironment{definition}[1][Definition]{\begin{trivlist}
\item[\hskip \labelsep {\bfseries #1}]}{\end{trivlist}}

\newcommand{\qed}{\nobreak \ifvmode \relax \else
      \ifdim\lastskip<1.5em \hskip-\lastskip
      \hskip1.5em plus0em minus0.5em \fi \nobreak
      \vrule height0.75em width0.5em depth0.25em\fi}

\newcommand{\beq}{\begin{equation}}
\newcommand{\eeq}{\end{equation}}
\newcommand{\barr}{\begin{array}}
\newcommand{\earr}{\end{array}}

\newcommand{\benum}{\begin{enumerate}}
\newcommand{\eenum}{\end{enumerate}}

\newcommand{\bit}{\begin{itemize}}
\newcommand{\eit}{\end{itemize}}

\newcommand{\bc}{\begin{center}}
\newcommand{\ec}{\end{center}}

\newcommand{\bdes}{\begin{description}}
\newcommand{\edes}{\end{description}}

\newcommand{\bfig}{\begin{figure}}
\newcommand{\efig}{\end{figure}}

\newcommand{\bemq}{\begin{quote} \begin{em}}
\newcommand{\eemq}{\end{em} \end{quote}}

\newcommand{\bmp}{\begin{minipage}}
\newcommand{\emp}{\end{minipage}}

\newcommand{\bsp}{\begin{slide*}}
\newcommand{\esp}{\end{slide*}}
\newcommand{\bsl}{\begin{slide}}
\newcommand{\esl}{\end{slide}}

\newcommand{\blem}{\begin{lemma}}
\newcommand{\elem}{\end{lemma}}
\newcommand{\bthm}{\begin{theorem}}
\newcommand{\ethm}{\end{theorem}}

\newcommand{\EX}[1]{\mathbb{E}\left[ #1 \right]} % expectation operator
\newcommand{\pr}[1]{\mathbb{P}\left[ #1 \right]}

\IEEEoverridecommandlockouts

\begin{document}

\title{Optimizing Age of Information in Wireless Networks with Perfect Channel State Information}
%\title{The Utility of Channel State Information in Optimizing Age of Information} %\title{Age-based Policies for Wireless Networks under General Interference Constraints}
%\date{12 April 2016}
\author{Rajat Talak, Sertac Karaman, and Eytan Modiano
\thanks{The authors are with the Laboratory for Information and Decision Systems (LIDS) at the Massachusetts Institute of Technology (MIT), Cambridge, MA. {\tt \{talak, sertac, modiano\}@mit.edu}}
\thanks{This work was supported by NSF Grants AST-1547331, CNS-1713725, and CNS-1701964, and by Army Research Office (ARO) grant number W911NF-17-1-0508.}
}

% \IEEEaftertitletext{\vspace{-0.6\baselineskip}}

\maketitle

\begin{abstract}
Age of information (AoI), defined as the time elapsed since the last received update was generated, is a newly proposed metric to measure the timeliness of information updates in a network. We consider AoI minimization problem for a network with general interference constraints, and time varying channels. We propose two policies, namely, virtual-queue based policy and age-based policy when the channel state is available to the network scheduler at each time step. We prove that the virtual-queue based policy is nearly optimal, up to a constant additive factor, and the age-based policy is at-most factor $4$ away from optimality.
%Numerical simulations suggest that the two policies are, in fact, very close to optimality.
Comparing with our previous work, which derived age optimal policies when channel state information is not available to the scheduler, we demonstrate a $4$ fold improvement in age due to the availability of channel state information.
\end{abstract}

\section{Introduction}
\label{sec:intro}
Timely delivery of information updates is gaining increasing relevance with the advent of technologies such as cyber-physical systems, internet of things, and futuristic unmanned aerial vehicular networks. In unmanned aerial vehicular networks, timely delivery of status updates, such as vehicle position and velocity, may be critical to network safety~\cite{FANETs2014, talakCDC16}. In internet of things or cyber-physical systems, timely delivery of sensor information can significantly improve the overall system performance~\cite{KimKumar2012_CPSperspective}.

\begin{figure}
  \centering
  \includegraphics[width=0.4\textwidth]{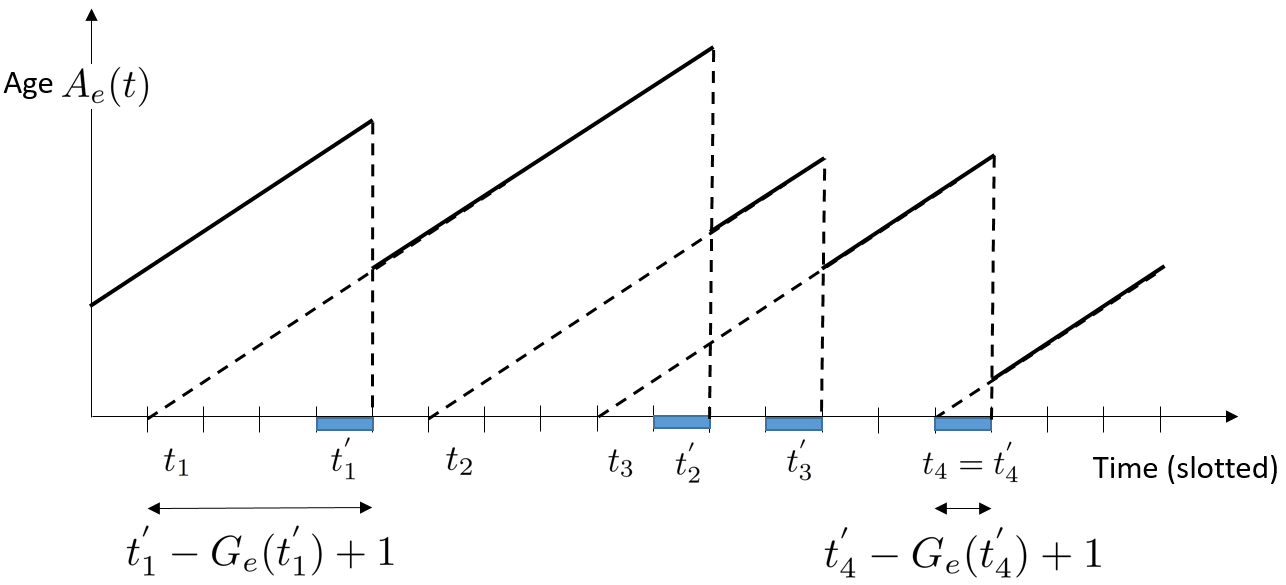} % 0.5 for single column
  \caption{Time evolution of age, $A_{e}(t)$, of a link $e$. Times $t_i$ and $t^{'}_i$ are instances of $i$th packet generation and reception, respectively. Given the definition $G_{e}(t^{'}_{i}) \triangleq t_{i}$, the age is reset to $t^{'}_i - G_{e}(t^{'}_{i}) + 1$ when the $i$th packet is received.}
  \label{fig:age}
\end{figure}

Age of information (AoI) is a recently proposed metric that measures the time elapsed since the last received update was generated by the source~\cite{2011SeCON_Kaul, 2012Infocom_KaulYates}. Figure~\ref{fig:age} shows typical evolution of AoI at a destination node, as a function of time. AoI, upon reception of a new update packet drops to the time elapsed since generation of the packet, and grows linearly otherwise. Therefore, AoI is a destination node centric measure, unlike packet delay, and is more suited for applications involving dissemination of time sensitive information.

In~\cite{2011SeCON_Kaul}, a simulation study considered a network of vehicles exchanging status updates, with an 802.11 based communication infrastructure, and showed that the AoI is minimized at a certain optimal packet generation rate. It further showed that AoI can be improved by changing the queue discipline of the MAC layer FIFO queue to last-in-first-out (LIFO). This observation was theoretically proved under a general network setting in~\cite{BedewyISIT17_LIFO_opt}.
Motivated by~\cite{2011SeCON_Kaul}, AoI was analyzed for several queueing models~\cite{2012Infocom_KaulYates, 2015ISIT_LongBoEM, 2013ISIT_KamKomEp, 2016X_LongBo, 2012CISS_KaulYates, 2016ISIT_Najm, 2014ISIT_CostaEp}.

However, age minimization for a network under general interference constraints and channel uncertainty has received very little attention. A problem of scheduling finitely many update packets under physical interference constraints was shown to be a NP-hard problem in~\cite{2016Ep_WiOpt}. Age for a broadcast network, where only a single link can be activated at any time, was studied in~\cite{2016allerton_IgorAge, 2017ISIT_YuPin}. Preliminary analysis of age for a slotted ALOHA like random access was done in~\cite{2017X_KaulYates_AoI_ALOHA}, and a distributed algorithm for age optimal ALOHA was only recently proposed in~\cite{talak17_greece}. Age in multi-hope interference networks has been studied in~\cite{talak17_allerton}.

We considered the problem of age minimization for a wireless network under general interference constraints, and time varying channel, in~\cite{talak17_StGenIC_TechRep}. We considered two types of sources: \emph{active sources}, which generate fresh information in every slot, and \emph{buffered sources}, which cannot generate fresh information in every slot. We showed that for a network with active sources, a stationary scheduling policy, which schedule links according to a stationary probability distribution, is peak age optimal and factor-$2$ average age optimal. We also showed that the same scheduling policy, with a certain packet generation rate control, is nearly optimal in the buffered case.

In~\cite{talak17_StGenIC_TechRep}, however, the space of policies was limited to not using the channel state information. In this paper, we relax this assumption and consider scheduling policies which have perfect channel state information $\mathbf{S}(t)$ at every time slot $t$.
We limit ourselves to the active sources case, and propose two policies: virtual-queue based policy and age-based policy, which uses the current channel state information to make scheduling decisions. We show, via numerical simulations, that availability of channel state information can significantly improve the AoI performance of the network.

We prove that the virtual-queue based policy is nearly peak age optimal, up to an additive factor, while the age-based policy is at most a factor $4$ away from the optimal peak and average age. Similar result has been recently derived for another age-based policy proposed for broadcast network, in which only a single link can be activated, in~\cite{Igor18_infocom}. Numerical simulations suggest that this bound is pessimistic, and that the proposed scheme performs much better.%, and that the proposed policies are in fact nearly optimal.

In numerical simulations, we observe the benefit/utility of using channel state information in scheduling to minimize age, especially when the network has `high' level of interference or `bad' channel quality. We demonstrate by considering a specific network example that the gap in age performance between known channel state and unknown channel state can be as large as $4$ fold. Even though channel state information may not be perfectly available in certain network settings, this work establishes the utility of acquiring such channel state information for scheduling to minimize age. To the best of our knowledge, this is the first work to make this observation.

\section{System Model}
\label{sec:model}
Consider a wireless network $G = (V, E)$, where $V$ denotes the set of nodes and $E$ the set of directed links. Not all links can be activated simultaneously. Thus, we call a set $m \subset E$ that can be activated simultaneously without interference as a \emph{feasible activation set}. We use $\mathcal{A}$ to denote the collection of all feasible activation sets. We consider a slotted time system, where the slot duration is normalized to unity.

We use $S_{e}(t)$ to denote the channel process, where $S_{e}(t) = 1$ if the channel is in ON state at time $t$ and $S_{e}(t) = 0$ if the channel is in OFF state at time $t$. The space of all channel states is given by $\mathbb{S} = \{0, 1\}^{|E|}$.
We consider $\{ S_{e}(t)\}_{t \geq 0}$ to be independent and identically distributed (i.i.d.) across time $t$, with $\gamma_e = \pr{S_{e}(t) = 1} > 0$, for all $e \in E$. We call this the \emph{i.i.d. channel process}. Note that the channel process is not identically distributed across links, and that $\gamma_e$ can be different for different links $e \in E$.

We use $U_{e}(t)$ to denote transmission decision on link $e$ at time $t$. $U_{e}(t) = 1$ if link $e$ is scheduled to transmit at time $t$. Not all transmissions succeed even if the set of activated links is a feasible activation set due to channel uncertainties. A successful transmission occurs over link $e$, at time $t$, if and only if $U_{e}(t)S_{e}(t) = 1$.

\begin{figure}
  \centering
  \includegraphics[width=0.95\linewidth]{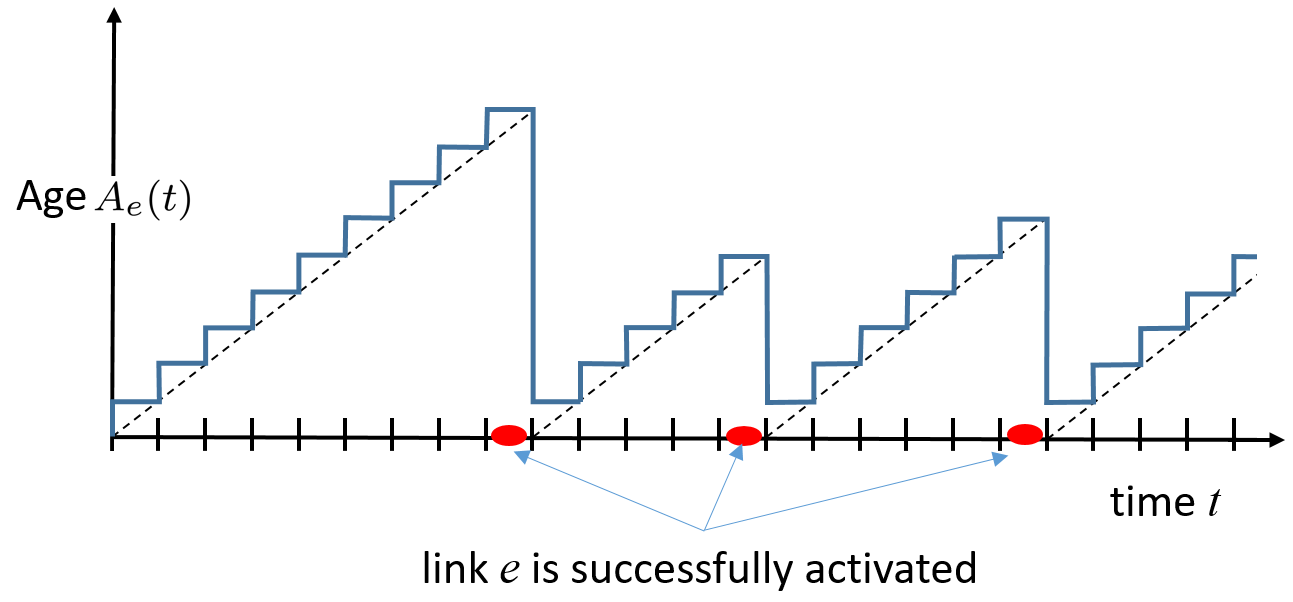}
  \caption{Evolution of age of link $e$, namely $A_{e}(t)$, as a function of time $t$.}\label{fig:age_evol}
\end{figure}

We consider \emph{active nodes}, which transmit fresh information at every transmission opportunity. We define age $A_{e}(t)$, of a link $e$ at time $t$, to be the time elapsed since the last successful activation of link $e$. Figure~\ref{fig:age_evol} shows evolution of age $A_{e}(t)$ for a link $e$. Age $A_{e}(t)$ reduces to $1$ upon a successful activation of link $e$, while it increases by $1$ in every slot in which there is no successful activation of link $e$, i.e.,
\begin{equation}\label{eq:age_evol}
A_{e}(t+1) = \left\{ \begin{array}{ll}
                       A_{e}(t) + 1 &\text{if}~U_{e}(t)S_{e}(t) = 0 \\
                       1 &\text{if}~U_{e}(t)S_{e}(t) = 1
                     \end{array} \right..
\end{equation}
This age evolution equation can more compactly be written as
\begin{equation}\label{eq:age_evol2}
A_{e}(t+1) = 1 + A_{e}(t) - U_{e}(t)S_{e}(t)A_{e}(t),
\end{equation}
for all $t \geq 0$ and $e \in E$.

We consider two popular age measures, namely, average age and peak age. Average age is the area under the age curve in Figure~\ref{fig:age_evol}, while peak age is the average of all the peaks of the age curve. More precisely, we define average age of a link $e$ as
\begin{equation}
\overline{A}^{\text{ave}}_{e} = \limsup_{t \rightarrow \infty} \EX{\frac{1}{t}\sum_{\tau = 0}^{t-1}A_{e}(\tau)},
\end{equation}
and the average age of the network to be the weighted sum
\begin{equation}
\overline{A}^{\text{ave}} = \sum_{e \in E}w_e \overline{A}^{\text{ave}}_{e}.
\end{equation}
Note that the sum of all the peaks, until time $t$, in the age curve can be expressed as $\sum_{\tau = 0}^{t}U_{e}(\tau)S_{e}(\tau)A_{e}(\tau)$. This is because $U_{e}(\tau)S_{e}(\tau) = 1$ only at times when age peaks. We, therefore, define the peak age to be
\begin{equation}
\label{eq:peak_def}
\overline{A}^{\text{p}}_{e} = \limsup_{t \rightarrow \infty} \frac{ \EX{\sum_{\tau = 0}^{t-1} U_{e}(\tau)S_{e}(\tau)A_{e}(\tau)} }{ \EX{\sum_{\tau = 0}^{t-1}U_{e}(\tau)S_{e}(\tau)} },
\end{equation}
for every link $e \in E$, and the peak age of the network to be the weighted sum
\begin{equation}
\overline{A}^{\text{p}} = \sum_{e \in E}w_e \overline{A}^{\text{p}}_{e}.
\end{equation}
We are interested in designing policies that minimize peak and average age.

Since both peak and average age are time average measures, performance of a policy $\pi$ does not depend on the initial age at time $0$. We, therefore, assume that the system starts with $A_{e}(0) = 0$ for all $e \in E$, unless stated otherwise.

%\rt{May not need this:}
%We could have defined, both peak and average age, without the expected value, namely, as
%\begin{equation}
%\label{eq:sp_Aave}
%A^{\text{ave}}_{e} = \limsup_{t \rightarrow \infty} \frac{1}{t}\sum_{\tau = 0}^{t-1}A_{e}(\tau),
%\end{equation}
%and
%\begin{equation}
%\label{eq:sp_Ap}
%A^{\text{p}}_{e} = \limsup_{t \rightarrow \infty} \frac{ \sum_{\tau = 0}^{t-1} U_{e}(\tau)S_{e}(\tau)A_{e}(\tau) }{ \sum_{\tau = 0}^{t-1}U_{e}(\tau)S_{e}(\tau) },
%\end{equation}
%whenever the respective limits exist in an almost sure sense. We show that for our proposed policies $A^{\text{ave}}_{e}$ and $A^{\text{p}}_{e}$ are also well defined, and $\overline{A}^{\text{ave}} = A^{\text{ave}}$ and $\overline{A}^{\text{p}} = A^{\text{p}}$, where $A^{\text{ave}} = \sum_{e \in E} w_e A^{\text{ave}}_{e}$ and $A^{\text{p}} = \sum_{e \in E} w_e A^{\text{p}}_{e}$.

\subsection{Unknown Channel Case}
\label{sec:unknown_channel}
In~\cite{talak17_StGenIC_TechRep}, we considered age minimization under the unknown channel case. Specifically, we considered all policies which scheduled feasible activation set $m_t \in \mathcal{A}$ at time $t$ as a function of the history
\begin{equation}
\mathcal{\hat{H}}(t) = \{ \mathbf{U}(\tau), \mathbf{A}(\tau') \big| 0 \leq \tau < t~\text{and}~0 \leq \tau' \leq t\}.
\end{equation}
We showed in~\cite{talak17_StGenIC_TechRep} that stationary policies, which schedule links according to a probability distribution that is independent of $\mathcal{\hat{H}}(t)$, is in fact peak age optimal and factor-$2$ average age optimal.

In stationary scheduling policies, every feasible activation set $m \in \mathcal{A}$ is assigned a fixed probability $x_m$ with which it is activated in slot $t$, independent across slots. The probability that a link $e \in E$ is activated in a slot is given by
\begin{equation}
f_e = \sum_{m: e \in m} x_m,
\end{equation}
for all $e \in E$. These set of equations can be compactly written as $\mathbf{f} = M\mathbf{x}$. Note that an activated link may fail in successfully transmitting the packet due to channel errors. The probability of successful activation of a link $e$ in any slot is $\alpha_e = \gamma_e f_e$, since the scheduling decision is independent of the current channel state.

Further, notice that, if a link $e$ is successfully activated with probability $\alpha_e = \gamma_ef_e$ in each slot, independent across slots, then the time since last transmission, i.e. age $A_{e}(t)$, is geometrically distributed with rate $\frac{1}{\gamma_e f_e}$. In~\cite{talak17_StGenIC_TechRep}, we showed that this is indeed equal to the peak age of link $e$, under stationary policy. As a result, the peak age for the stationary policy, determined by distribution $\mathbf{x}$, is given by $\overline{A}^{\text{p}} = \sum_{e \in E} \frac{w_e}{\gamma_e f_e}$, and thus, the optimal peak age is given by
\begin{align}
\label{eq:ex0}
\begin{aligned}
\overline{A}^{\text{p}\ast} =&~\underset{\mathbf{x}, \mathbf{f}}{\text{Minimize}} & & \sum_{e \in E} \frac{w_e}{\gamma_e f_e}, \\
&~\text{subject to} && \mathbf{f} = M\mathbf{x}, \\
&~&& \mathbf{1}^{T}\mathbf{x} \leq 1~\text{and}~\mathbf{x} \geq 0.
\end{aligned}
\end{align}
The peak age optimal stationary policy is obtained by solving~\eqref{eq:ex0}.

In the next sub-section, we discuss the space of policies considered in this paper, and show how knowing the channel state affects age minimization. We argue that in the case when channel state information is available for scheduling, age better than what is given by~\eqref{eq:ex0} can be achieved.

\subsection{Scheduling Policies}
\label{sec:scheduling_policies}
A scheduling policy determines the set of links $m_t \subset E$ that will be activated at each time $t$, i.e., $m_t = \{ e \in E | U_{e}(t) = 1 \}$. The policy can make use of the past history of link activations and observed channel states to make this decision, i.e., at each time $t$, the policy $\pi$ will determine $m_t$ as a function of the set
\begin{equation}
\mathcal{H}(t) = \{\mathbf{U}(\tau), \mathbf{S}(\tau'), \mathbf{A}(\tau')~|~0 \leq \tau < t,~0\leq \tau' \leq t \}.
\end{equation}
We consider centralized scheduling policies, in which this information is centrally available to a scheduler, which is also able to implement its scheduling decision. This assumption is consistent with that in network scheduling literature~\cite{TassiEp92, Neely_book}.

\begin{figure}
  \centering
  \includegraphics[width=0.90\linewidth]{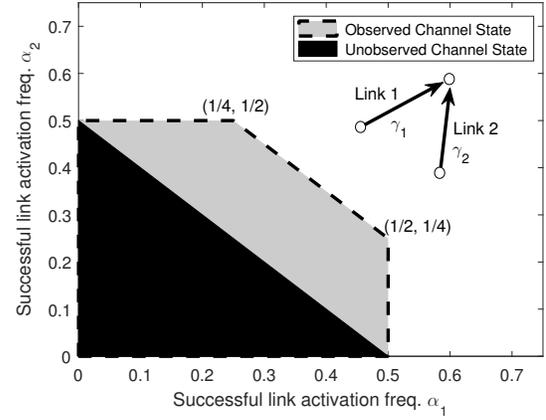}
  \caption{Plot of achievable successful link activation frequency regions for the two link network, in which only one link can be activated at a time. Shown are regions when channel state is observed (grey) and unobserved (black).}\label{fig:rate_region_plot}
\end{figure}
To see the difference between age minimization under known and unknown channel process consider the two link example shown in Figure~\ref{fig:rate_region_plot}. In this example, only one link can be activated at a time. Let the weights $w_1 = w_2 = 1$ for the two links, and the channel success probabilities be $\gamma_1 = \gamma_2 = 0.5$. When the channel state $\mathbf{S}(t) = (S_{1}(t), S_{2}(t))$ is unavailable the peak age minimization problem is given by (from~\eqref{eq:ex0}):
\begin{align}
\label{eq:ex1_a}
\begin{aligned}
\overline{A}^{\text{p}\ast} =&~\underset{f_1, f_2}{\text{Minimize}} & & \frac{1}{\gamma_1 f_1} + \frac{1}{\gamma_2 f_2}, \\
&~\text{subject to} && f_1 + f_2 \leq 1, \\
&~&& f_1 \geq 0~\text{and}~f_2 \geq 0.
\end{aligned}
\end{align}
Here, $f_1$ denotes the fraction of times link $1$ is scheduled and $f_2$ denotes the fraction of times link $2$ is scheduled. Since $\gamma_1 = \gamma_2 = 0.5$, the optimal solution to~\eqref{eq:ex1_a} is given by $f^{\ast}_{1} = f^{\ast}_{2} = 0.5$, i.e. with probability $0.5$ each link gets scheduled in each slot, and as a result the optimal peak age is $\overline{A}^{\text{p}\ast} = 8$.

However, if we can observe the channel state $\mathbf{S}(t)$ in every slot before making scheduling decision, we can achieve even smaller age than $\overline{A}^{\text{p}\ast} = 8$. Consider the following policy: schedule link $1$ whenever $S_{1}(t) = 1$, else schedule link $2$. The successful link activation frequency on link $1$ is then $\alpha_1 = \gamma_1 = 0.5$, while on link $2$ it is $\alpha_2 = \gamma_2(1-\gamma_1) = 0.25$. The peak age is given by $\overline{A}^{\text{p}} = \frac{1}{\alpha_1} + \frac{1}{\alpha_2} = 6 < \overline{A}^{\text{p}\ast} = 8$. This happens primarily because the set of achievable successful link activation frequencies, namely $\alpha_e$, is larger in the case when the channel can be observed before deciding on the schedule in each slot. In Figure~\ref{fig:rate_region_plot}, we show these regions in the observed and unobserved channel state case for the two link example.

This shows that when the channel state is available for making scheduling decisions, the network age performance can be improved upon. In the next sub-section we define a sub-class of policies that make scheduling decision based only on the current channel state $\mathbf{S}(t)$, and not the entire history $\mathcal{H}(t)$. We will see later that these policies can be peak age optimal.

\subsection{$\mathcal{S}$-only policies}
Just as the stationary policies turn out to be peak age optimal in the unknown channel case, we define a sub-class of policies that are peak age optimal in the known channel case. These policies do not use any past history, but only the current channel state $\mathbf{S}(t)$, defined as follows~\cite{Neely_book}:
\begin{framed}
\begin{definition}[$\mathcal{S}$-only policy:]
For each observed channel state $S \in \mathbb{S}$ we assign a probability distribution $p(S,m)$ over the set of feasible activation sets $m \in \mathcal{A}$. If channel state $\mathbf{S}(t)$ is observed then the activation set $m \in \mathcal{A}$ is activated for slot $t$ with probability $p\left(\mathbf{S}(t), m\right)$.
\end{definition}
\end{framed}
For an $\mathcal{S}$-only policy, the rate at which a successful transmission occurs over link $e$ is given by
\begin{align}
\alpha_e = \EX{U_{e}(t)S_{e}(t)} &= \pr{U_{e}(t)S_{e}(t) = 1}, \nonumber \\
&= \gamma_e \pr{U_{e}(t) = 1 | S_{e}(t) = 1},
\end{align}
for all $e \in E$. The space of all such rates $\bm{\alpha}$ will depend on channel success probabilities $\gamma_e$, and thus, we use $\Lambda_{\mathcal{S}}(\bm{\gamma})$ to denote this space of all feasible $\bm{\alpha}$ using $\mathcal{S}$-only policy. For the two link example in Figure~\ref{fig:rate_region_plot}, $\Lambda_{\mathcal{S}}(\bm{\gamma})$ is exactly the grey region of successful link activation frequencies $(\alpha_1, \alpha_2)$.
It is known that if $\Lambda(\bm{\gamma})$ is the space of rates $\bm{\alpha}$ achievable under all policies then $\Lambda(\bm{\gamma}) = \Lambda_{\mathcal{S}}(\bm{\gamma})$~\cite{Neely_book}. This will help us show that an $\mathcal{S}$-only policy is peak age optimal.

\section{Problem Formulation}
\label{sec:formulation}
In this section, we formulate the peak and average age minimization problems under a general channel process. To do so in a meaningful way, we would like to restrict our search to a certain reasonable policy space. We consider the following policy spaces:
\begin{equation}\nonumber
\overline{\Pi}_1 = \left\{ \pi \Big|~\exists B~\text{s.t.}~\EX{A^{\pi}_{e}(t)} \leq B~\forall~t \geq 0~\text{and}~e \in E\right\},
\end{equation}
and
\begin{equation}\nonumber
\overline{\Pi}_2 = \left\{ \pi \Big|~\exists B~\text{s.t.}~\EX{A^{\pi 2}_{e}(t)} \leq B~\forall~t \geq 0~\text{and}~e \in E\right\}.
\end{equation}
Firstly, note that the constraints that the first and second moment of age $A_{e}(t)$ should not grow in $t$ is natural, because $A_{e}(t)$ is the time since last successful transmission on link $e$. It growing in time would necessarily mean that the transmissions are becoming less frequent as time goes by.

We consider the policy space $\overline{\Pi}_1$ for peak age minimization, while space $\overline{\Pi}_2$ for average age minimization. For a `good' policy, we anticipate the process $\{ \mathbf{A}(t) \}_{t}$ to be ergodic, in which case the policy is in $\overline{\Pi}_1$. For `good' average age policy, it stands to reason that ergodicity of $\{ \mathbf{A}^{2}(t) \}_{t}$ would be required. This is because the average age, being the area under the age curve, depends on $A^{2}(t)$.

We define optimal peak and average age to be
\begin{equation}
\overline{A}^{\text{p}\ast} = \min_{\pi \in \overline{\Pi}_1} \overline{A}^{\text{p}}(\pi)~~~~\text{and}~~~~\overline{A}^{\text{ave}\ast} = \min_{\pi \in \overline{\Pi}_2} \overline{A}^{\text{ave}}(\pi),
\end{equation}
where the minimization is over the space $\overline{\Pi}_1$ for peak age and over $\overline{\Pi}_2$ for average age. Note that $\overline{\Pi}_2 \subset \overline{\Pi}_1$ since $\EX{A_{e}(t)} \leq \sqrt{\EX{A^{2}_{e}(t)}}$ by Jensen's inequality.

\subsection{Peak Age Minimization}
\label{sec:peak_age_form}
We first present a lemma that states a conservation law for age. Intuitively, it states that for any policy $\pi \in \overline{\Pi}_1$, the sum of all age peaks is equal to the total time elapsed plus a small insignificant term that goes to $0$ as $t \rightarrow \infty$.
\begin{framed}
\begin{lemma}
\label{lem:time_conservation}
For any policy $\pi \in \overline{\Pi}_1$ we have
\begin{equation}
\lim_{t \rightarrow \infty} \EX{\frac{1}{t}\sum_{\tau = 0}^{t-1}U_{e}(\tau)S_{e}(\tau)A_{e}(\tau)} = 1,
\end{equation}
for all $e \in E$.
\end{lemma}
\end{framed}
\begin{IEEEproof}
See Appendix~\ref{pf:lem:time_conservation}.
\end{IEEEproof}

A direct consequence of Lemma~\ref{lem:time_conservation} is that the peak age minimization problem $\min_{\pi \in \overline{\Pi}_1} \overline{A}^{\text{p}}(\pi)$ reduces to
\begin{align}
\label{eq:peak_min_problem}
\begin{aligned}
& \underset{\bm{\alpha} \geq 0, \pi \in \overline{\Pi}_1}{\text{Minimize}} & & \sum_{e \in E}\frac{w_e}{\alpha_e}, \\
& \text{subject to} && \liminf_{t \rightarrow \infty} \EX{\frac{1}{t}\sum_{\tau = 0}^{t-1}U_{e}(\tau)S_{e}(\tau)} \geq \alpha_e~~\forall~e \in E.
\end{aligned}
\end{align}
We prove this equivalence in Appendix~\ref{pf:der_peak_min_prob}. This result is significant because it shows that the peak age minimization problem is independent of the age evolution equation. For this reason peak age minimization is much simpler than average age minimization. %Note that~\eqref{eq:peak_min_problem} is valid with no assumptions on the channel process $\{ S_{e}(t)\}_{t \geq 0}$.
We propose a virtual-queue based algorithm in Section~\ref{sec:virtualQ-policy} to solve this problem. %, and prove that it is near optimal when the channel process is i.i.d..

\subsection{Average Age Minimization}
\label{sec:ave_age_form}
In this section, we provide an equivalent formulation for average age minimization under general channel process. By definition, we know that the average age for a link $e$ is given by
\begin{equation}
\overline{A}^{\text{ave}}_{e} = \limsup_{t \rightarrow \infty} \EX{\frac{1}{t}\sum_{t=0}^{t-1}A_{e}(\tau)}.
\end{equation}
The following result provides a different characterization of the average age in terms of $A_{e}^{2}(t)$, for all $\pi \in \overline{\Pi}_2$. This result will be useful to get an intuitive grasp over the class of policies proposed in Section~\ref{sec:age-square-policy}.
\begin{framed}
\begin{lemma}
\label{lem:ave_age_conservation}
Define $B_{e}(t) = A^{2}_{e}(t) + \beta A_{e}(t)$ for all $t$ and $e \in E$, and any given $\beta \in \mathbb{R}$. Then, for $\pi \in \overline{\Pi}_2$, we have
\begin{multline}
\overline{A}^{\text{ave}}_{e} = \frac{1}{2}\limsup_{t \rightarrow \infty} \EX{\frac{1}{t}\sum_{\tau = 0}^{t-1} U_{e}(\tau)S_{e}(\tau) B_{e}(\tau)} + \frac{1-\beta}{2}, \nonumber
\end{multline}
for all $e \in E$.
\end{lemma}
\end{framed}
\begin{IEEEproof}
See Appendix~\ref{pf:lem:ave_age_conservation}.
\end{IEEEproof}
For an intuitive understanding of Lemma~\ref{lem:ave_age_conservation}, note that average age is essentially the average area of the triangles formed by the age curve in Figure~\ref{fig:age_evol}. Note that $S_{e}(t)U_{e}(t)A^{2}_{e}(t)$ are square of age peaks in Figure~\ref{fig:age_evol}, because $S_{e}(t)U_{e}(t) = 1$ only at the instances when there is a successful transmission on link $e$. The additional term of $\beta A_{e}(t)$ is due to Lemma~\ref{lem:time_conservation}.

Lemma~\ref{lem:ave_age_conservation} also implies that average age minimization problem over $\pi \in \overline{\Pi}_2$ can be equivalently posed to minimize
\begin{equation}\label{eq:ave_age_proxy}
\limsup_{t \rightarrow \infty} \EX{\frac{1}{t}\sum_{\tau = 0}^{t-1}\sum_{e \in E} w_e U_{e}(\tau)S_{e}(\tau)B_{e}(\tau)},
\end{equation}
where $B_{e}(\tau) = A_{e}^{2}(\tau) + \beta A_{e}(\tau)$, for all $\tau \geq 0$, $e \in E$, and any chosen $\beta \in \mathbb{R}$. Since, age reduces to $1$ after a link activation it makes intuitive sense to choose $\mathbf{U}(t)$ such that as
\begin{equation}
\label{eq:ave_age_decision}
\mathbf{U}(t) = \arg\max_{\mathbf{U}^{'}(t)} \sum_{e \in E} w_e U^{'}_{e}(t)S_{e}(t)\left[ A^{2}_{e}(t) + \beta A_{e}(t)\right],
\end{equation}
in time slot $t$. This, in the least, should minimize age in the next slot. We analyze such policies in Section~\ref{sec:age-square-policy}, and show that these policies are within a factor of $4$ away from the optimal average age $\overline{A}^{\text{ave}\ast}$. However, in simulations we observe that these policies are very close to optimal.

\subsection{Bounds on Peak and Average Age}
\label{sec:age_bounds}
In this sub-section, we provide a characterization of optimal peak age $\overline{A}^{\text{p}\ast}$ and a lower-bound on average age.
We first characterize the optimal peak age by showing that a $\mathcal{S}$-only policy is peak age optimal.
\begin{framed}
\begin{theorem}
\label{thm:peak_age_lb}
The optimal peak age $\overline{A}^{\text{p}\ast}$ is given by
\begin{align}\label{eq:age_opt}
\begin{aligned}
\overline{A}^{\text{p}\ast} =&~~\underset{\bm{\alpha}}{\text{Minimize}}
& & \sum_{e \in E} \frac{w_{e}}{\alpha_e}, \\
&~~\text{subject to} && \bm{\alpha} \in \Lambda_{\mathcal{S}}\left( \bm{\gamma}\right),
\end{aligned}
\end{align}
and as a consequence, there exists a $\mathcal{S}$-only policy that minimizes peak age, and it can be obtain by solving~\eqref{eq:age_opt}.
\end{theorem}
\end{framed}
\begin{IEEEproof}
The optimality of $\mathcal{S}$-only policies in solving the problem~\eqref{eq:peak_min_problem} follows from Theorem~4.5 in~\cite{Neely_book}. In Appendix~\ref{pf:thm:peak_age_lb}, we show that the peak age minimization problem over the space of $\mathcal{S}$-only policies can be written as~\eqref{eq:age_opt}.
\end{IEEEproof}
Theorem~\ref{thm:peak_age_lb} can be used to obtain a peak age optimal $\mathcal{S}$-only policy. However, the search space $\Lambda_{\mathcal{S}}\left( \bm{\gamma}\right)$ is usually difficult to characterize for general interference constraints. Another issue is that, to solve~\eqref{eq:age_opt}, requires exact knowledge of the channel statistics $\gamma_e$. We propose two algorithms that attain near optimal peak and average age, even when the channel statistics $\gamma_e$ is not known apriori but learned on the fly by observing channel states $\mathbf{S}(t)$.

We now proceed to derive a lower-bound on average age.
\begin{framed}
\begin{lemma}
\label{lem:peak_ave_bound}
For any policy $\pi \in \overline{\Pi}_2$, we have
\begin{equation}
\overline{A}^{\text{p}}\left( \pi \right) \leq 2\overline{A}^{\text{ave}}\left( \pi \right) - \sum_{e \in E} w_e.
\end{equation}
And as a consequence the same relation also holds at optimality, namely, $\overline{A}^{\text{p}\ast} \leq 2\overline{A}^{\text{ave}\ast} - \sum_{e \in E}w_e$.
\end{lemma}
\end{framed}
\begin{IEEEproof}
See Appendix~\ref{pf:lem:peak_ave_bound}.
\end{IEEEproof}
Lemma~\ref{lem:peak_ave_bound} provides us with a natural lower-bound on the optimal average age $\overline{A}^{\text{ave}\ast}$ in terms of the optimal peak age. Since, the optimal peak age can be obtained from Theorem~\ref{thm:peak_age_lb} we get
\begin{equation}
\frac{1}{2}\sum_{e \in E}\frac{w_e}{\alpha^{\ast}_{e}} + \frac{1}{2}\sum_{e \in E}w_e \leq \overline{A}^{\text{ave}\ast},
\end{equation}
where $\bm{\alpha}^{\ast}$ is a solution to the optimization problem~\eqref{eq:age_opt}.

%\rt{Can we say that there exists a policy which is both peak and average age optimal? Perhaps average age optimality implies peak age optimality but not vice versa.}

%\rt{We showed that for iid channel model and stationary policy peak age equals average age. Is such a relation possible for stationary, ergodic channel and stationary policy?}

\section{Virtual-Queue Based Policy}
\label{sec:virtualQ-policy}
We now propose a policy that solves the peak age minimization problem~\eqref{eq:peak_min_problem}. Note that a policy $\pi$ can decide on the activation set $m_t$, at time $t$, based on the entire history $\mathcal{H}(t)$. However, we do not need the entire history to make a choice at time $t$ but only a representation of it.

To do so, we construct virtual queue $Q_{e}(t)$, which reduce by (at most) $1$ upon a successful transmission over link $e$ and increased otherwise. These queue lengths determine the `value' of scheduling link $e$ in time slot $t$. Therefore, a set $m_t \in \mathcal{A}$ that maximizes $\sum_{e \in m} w_e Q_{e}(t) S_{e}(t)$ is activated in slot $t$.
This virtual-queue based policy, call it $\pi_{Q}$, is described below. Here, $V > 0$ is any chosen constant.
\begin{framed}
\textbf{Age based policy $\pi_Q$}
Start with $Q_{e}(0) = 1$ for all $e \in E$. At time $t$,
\begin{enumerate}
  \item Schedule activation set $m_t$ given by
        \begin{equation}
        m_{t} = \arg\max_{m \in \mathcal{A}} \sum_{e \in m} w_e Q_{e}(t)S_{e}(t),
        \end{equation}
  \item Update $Q_{e}(t)$ as
        \begin{equation}
        Q_{e}(t+1) = \left[Q_{e}(t) + \sqrt{\frac{V}{Q_{e}(t)}} - U_{e}(t)S_{e}(t)\right]_{+1}, \nonumber
        \end{equation}
        for all $e \in E$, where $[x]_{+1} = \max\{x, 1\}$.
\end{enumerate}
\end{framed}

We now prove that the policy $\pi_Q$ is nearly peak age optimal up to an additive factor.
\begin{framed}
\begin{theorem}
\label{thm:piQ}
The peak age for policy $\pi_Q$ is bounded by
\begin{equation}
\label{eq:thm:piQ}
\overline{A}^{\text{p}}(\pi_{Q}) \leq \overline{A}^{\text{p}\ast} + \frac{1}{2}\sum_{e \in E} w_e + \frac{1}{2V}\sum_{e \in E} w_e,
\end{equation}
where $\overline{A}^{\text{p}\ast}$ is the optimal value of~\eqref{eq:age_opt}.
\end{theorem}
\end{framed}
\begin{IEEEproof}
Let $\alpha_{e}(t) = \sqrt{\frac{V}{Q_{e}(t)}}$ and $\overline{\alpha}_{e}(t) = \frac{1}{t}\sum_{\tau = 0}^{t-1}\alpha_{e}(\tau)$ for all $t \geq 0$ and $e \in E$. Also, let $g(\mathbf{\alpha}) = \sum_{e \in E} \frac{w_e}{\alpha_e}$ be the objective function in our optimization problem~\eqref{eq:peak_min_problem}.
The proof is divided into three parts:

\textbf{Part A}: For all time $t$, we have
        \begin{equation}\label{eq:DPP_piQ_opt}
        \limsup_{t \rightarrow \infty} \EX{g\left( \overline{\bm{\alpha}}(t)\right)} \leq \overline{A}^{\text{p}\ast} + \frac{1}{2}\sum_{e \in E}w_e + \frac{1}{2V}\sum_{e \in E}w_e.
        \end{equation}

\textbf{Part B}: The virtual queue $\mathbf{Q}(t)$ is mean rate stable, i.e., for all $e \in E$ we have
        \begin{equation}
        \label{eq:b1}
        \limsup_{t \rightarrow \infty} \frac{1}{t}\EX{Q_{e}(t)} = 0.
        \end{equation}

\textbf{Part C}: If $\mathbf{Q}(t)$ is mean rate stable then
        \begin{equation}
        \label{eq:c1}
        \liminf_{t \rightarrow \infty} \EX{\overline{\alpha}_{e}(t)} \leq \liminf_{t \rightarrow \infty} \frac{1}{t}\EX{\sum_{\tau=0}^{t-1}U_{e}(\tau)S_{e}(\tau)},
        \end{equation}
        and
        \begin{equation}
        \label{eq:c2}
        \overline{A}^{\text{p}}\left( \pi_Q\right) \leq \limsup_{t \rightarrow \infty} \EX{g\left( \overline{\bm{\alpha}}(t)\right)}.
        \end{equation}

The proofs of Part A, B, and C are given in Appendix~\ref{pf:thm:piQ}.
%
%Due to space constraints, the proofs of Part A, B, and C are given in our technical report~\cite{}.
%
Since the virtual queues are mean rate stable, by Part~B,~\eqref{eq:c1} and~\eqref{eq:c2} are true. From~\eqref{eq:DPP_piQ_opt} and~\eqref{eq:c2} we get the result in~\eqref{eq:thm:piQ}.

Further, if we set
\begin{equation}
\alpha^{V}_{e} = \liminf_{t \rightarrow \infty} \EX{\overline{\alpha}_{e}(t)},
\end{equation}
for each $e \in E$, then $\bm{\alpha}^{V}$, with policy $\pi_Q$, solves the optimization problem~\eqref{eq:peak_min_problem}, up to an additive factor. To see this, notice that from~\eqref{eq:c1}, we know that $\bm{\alpha}^{V}$ satisfies the inequality constraint in~\eqref{eq:peak_min_problem}. Now, consider the objective function evaluated at $\bm{\alpha}^{V}$:
\begin{align}
g\left( \bm{\alpha}^{V} \right) &= g\left( \liminf_{t \rightarrow \infty} \EX{\overline{\bm{\alpha}}(t)} \right), \nonumber \\
&= \limsup_{t \rightarrow \infty} g\left(\EX{\overline{\bm{\alpha}}(t)} \right), \nonumber \\
&\leq \limsup_{t \rightarrow \infty} \EX{ g\left( \overline{\bm{\alpha}}(t) \right) }, \label{eq:oyo}
\end{align}
where the first equality is because $g$ is a continuous decreasing function in $\bm{\alpha}$, while the second inequality follows directly from Jensen's inequality as $g$ is convex.
Substituting~\eqref{eq:DPP_piQ_opt} in~\eqref{eq:oyo} we get
\begin{equation}
g\left( \bm{\alpha}^{V} \right) \leq \overline{A}^{\text{p}\ast} + \frac{1}{2}\sum_{e \in E}w_e + \frac{1}{2V}\sum_{e \in E}w_e.
\end{equation}
\end{IEEEproof}

%\rt{Give an extension, where $S_{e}(t)$ cannot be observed for every slot, but an estimate $\hat{\gamma}_e$ is obtained based on past transmission successes and failures.}
Theorem~\ref{thm:piQ} shows that even when the channel statistics are not known the optimal peak age $A^{\text{p}\ast}$ can be achieved, up to an additive factor of $\frac{1}{2}\sum_{e \in E}w_e$, with arbitrary precision. The precision can be chosen by selecting $V$. For example, we may obtain peak age of at most $\overline{A}^{\text{p}\ast} + \frac{1}{2}\sum_{e \in E}w_e + \epsilon$ by setting $V = \frac{1}{2\epsilon}\sum_{e \in E}w_e$. %However, just as in Lypunov optimization~\cite{Neely_book} there is a cost to having a larger $V$, which is larger virtual queue lengths.
%

%However, as is known~\cite{Neely_book}, $V$ also impacts convergence time of the policy $\pi_Q$. The convergence time for $\pi_Q$ is $O(\sqrt{V})$. Choice of $V$ has to be made by considering this convergence time vs optimality tradeoff.

%\rt{Add result about convergence of $A^{\text{p}}$}.

\section{Age-Based Policy}
\label{sec:age-square-policy}
We now consider an age based policy, which schedule links as a function of links' age $A_{e}(t)$. Lemma~\ref{lem:ave_age_conservation} provided an alternate characterization of average age $A^{\text{ave}}$ under policy $\pi \in \overline{\Pi}_2$ given by
\begin{multline}
A^{\text{ave}}(\pi) = \limsup_{t \rightarrow \infty} \frac{1}{t}\EX{\sum_{\tau=0}^{t-1} \sum_{e \in E} w_e U_{e}(\tau)S_{e}(\tau)B_{e}(\tau)} \\
+ \frac{1-\beta}{2}\sum_{e \in E}w_e, \nonumber
\end{multline}
where $B_{e}(t) = A_{e}^{2}(t) + \beta A_{e}(t)$ and $\beta \in \mathbb{R}$. We now propose an age-based policy that schedules set $m_t \in \mathcal{A}$ with maximum weight $\sum_{e \in m} w_e S_{e}(t)\left[ A^{2}_{e}(t) + \beta A_{e}(t)\right]$.
\begin{framed}
\textbf{Age-based Policy $\pi_{A}$}
The policy activates links $m_{t} \in \mathcal{A}$ in slot $t$ given by: \begin{equation}
m_{t} = \arg\max_{m \in \mathcal{A}} \sum_{e \in m} w_e S_{e}(t)\left[A^{2}_{e}(t) + \beta A_{e}(t)\right],
\end{equation}
for all $t \geq 1$.
\end{framed}
We, now, prove bounds on the peak and average age of policy $\pi_A$. More specifically, we show that the average and peak age of policy $\pi_A$ is within a factor of $4$ from the respective optimal values.
\begin{framed}
\begin{theorem}
\label{thm:piA}
The age-based policy policy $\pi_A$ is at most factor-4 peak and average age optimal, i.e.,
\begin{equation}
\label{eq:thm:piA_1}
\overline{A}^{\text{ave}}(\pi_A) \leq 4\overline{A}^{\text{ave}\ast} - c_1(\beta)\sum_{e \in E} w_e,
\end{equation}
and
\begin{equation}
\label{eq:thm:piA_2}
\overline{A}^{\text{p}}(\pi_A) \leq 4\overline{A}^{\text{p}\ast} - c_2(\beta)\sum_{e \in E}w_e,
\end{equation}
where $c_1(\beta) = \frac{10 + 2\beta - \beta^2}{4}$ and $c_2(\beta) = \frac{4 + 2\beta - \beta^2}{2}$.
\end{theorem}
\end{framed}
\begin{IEEEproof}
To obtain the bound we define two functions $f(t)$ and $\Delta(t)$, where $f(t)$ is a representation of our objective function, which is age at time $t$, while $\Delta(t)$ is the drift of a certain Lyapunov function $L(t)$. We obtain a bound on $\EX{f(t) + \Delta(t)|\mathbf{A}(t)}$ where $\mathbf{A}(t)$ denote the vector of all $A_{e}(t)$. Telescoping $f(t) + \Delta(t)$ over $T$ slots then yields the result. The detailed proof is given in Appendix~\ref{pf:thm:piA}.
%The detailed proof is given in our technical report~\cite{}.
\end{IEEEproof}
We note that $\beta$ can be chosen to improve the additive factor of optimality. The best bounds, for both peak and average age, occur when $\beta = 1$, for which both $c_1(\beta)$ and $c_{2}(\beta)$ are maximized.
In the next section, we evaluate the age-based policies for different choices of $\beta$. We also compare it with the virtual-queue based policy $\pi_Q$ from Section~\ref{sec:virtualQ-policy}.

\section{Numerical Results}
\label{sec:numerical}
Consider a network of $N = 20$ links, in which at most $K$ links can be activated at any given time. We numerically study the performance of our proposed scheduling policies for this network. We set $w_e = 1$ for all links $e$. We assume links to be either `good', in which case $\gamma_e = \gamma_{\text{good}} = 0.9$, or `bad' in which case $\gamma_e = \gamma_{\text{bad}} = 0.1$. We use $n_{\text{bad}}$ to denote the number of bad links in the network. We simulate the policies $\pi_Q$, $\pi_A$, and the optimal policy for the unknown channel case, proposed in~\cite{talak17_StGenIC_TechRep}, over $10^5$ time slots.

\begin{figure}
  \centering
  \includegraphics[width=0.83\linewidth]{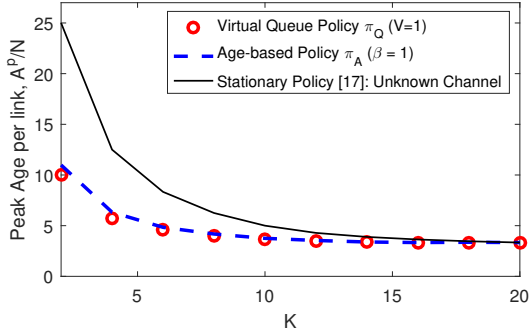}
  \caption{Per-link peak age, $A^{\text{p}}/N$, for various policies as a function of $K$.}\label{fig:main7peak}
\end{figure}

\begin{figure}
  \centering
  \includegraphics[width=0.85\linewidth]{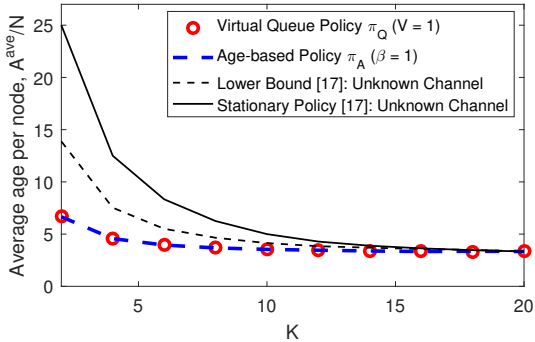}
  \caption{Per-link average age, $A^{\text{ave}}/N$, for various policies as a function of $K$.}\label{fig:main7ave}
\end{figure}

In Figure~\ref{fig:main7peak} and~\ref{fig:main7ave}, we plot per-link peak and average age, namely $A^{\text{p}}/N$ and $A^{\text{ave}}/N$, as a function of $K$. Here, we have chosen the parameters $V = 1$ for the virtual-queue policy $\pi_Q$, and $\beta = 1$ for the age-based policy $\pi_A$. We observe that the peak and average age of the virtual-queue based policy $\pi_Q$ and the age-based policy $\pi_A$ nearly coincide.

Also plotted in Figures~\ref{fig:main7peak} and~\ref{fig:main7ave}, is the case when the channel state is not observed, i.e., scheduling decisions are made only using history $\hat{\mathcal{H}}(t)$. We plot the peak age optimal policy $\pi_C$ of~\cite{talak17_StGenIC_TechRep}, while in Figure~\ref{fig:main7ave}, we also plot a lower-bound on average age that can be achieved by any such policy~\cite{talak17_StGenIC_TechRep}, since $\pi_C$ is not average age optimal.
We observe that the gap between the optimal policy $\pi_C$ in the unknown channel case and policies $\pi_Q$, $\pi_A$ of the known channel case is large when $K$ is small, and diminishes as $K$ increases.
Smaller $K$ implies more network interference, as fewer links can be activated simultaneously. This shows that there is a significant utility, in terms of age reduction, in knowing the channel state especially when the network suffers from large interference. %When $K=2$, for example, the age gap is as large as $7$ fold.

\begin{figure}
  \centering
  \includegraphics[width=0.85\linewidth]{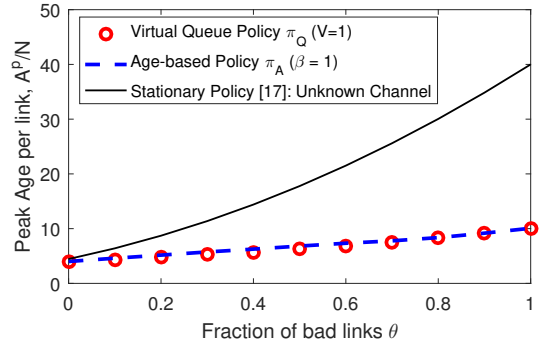}
  \caption{Per-link peak age, $A^{\text{p}}/N$, for various policies as a function of $\theta$.}\label{fig:main10peak}
\end{figure}

\begin{figure}
  \centering
  \includegraphics[width=0.85\linewidth]{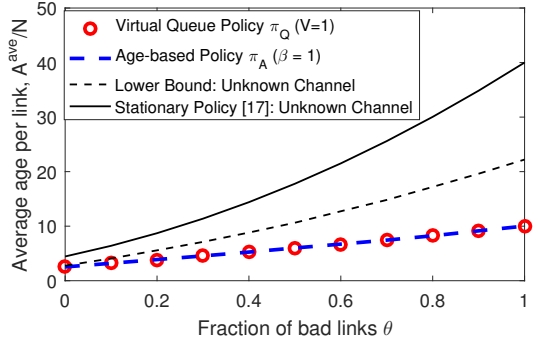}
  \caption{Per-link average age, $A^{\text{ave}}/N$, for various policies as a function of $\theta$.}\label{fig:main10ave}
\end{figure}

In Figure~\ref{fig:main10peak} and~\ref{fig:main10ave} we plot per-link peak and average age as a function of the fraction of nodes with bad channel, namely $\theta = \frac{n_{\text{bad}}}{N}$. We observe that the gap between the optimal policy $\pi_C$ in the unknown channel state case, and our policies $\pi_Q$ and $\pi_A$ of the known channel case, widens as the fraction $\theta$ increases. This indicates that if the channel statistics of the network are poor then there is a significant utility, in terms of age reduction, in knowing the channel state information. For example, when all channels are `bad', i.e. $\theta = 1$, the gap is as large as $4$ fold.

\subsection{Choice of Parameters $V$ and $\beta$}
\begin{figure}
  \centering
  \includegraphics[width=0.85\linewidth]{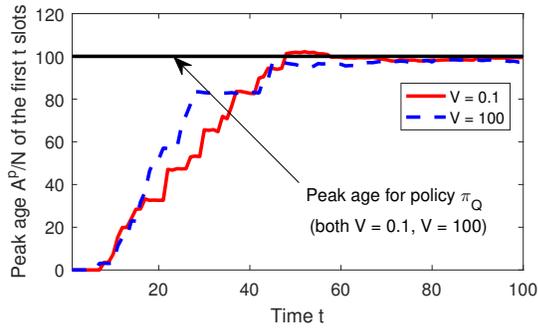}
  \caption{Per-link peak age, $A^{\text{p}}(\pi_Q)/N$, computed till time $t$ for the virtual-queue policy $\pi_Q$ for $V = 0.1$ and $V=100$. Also plotted is the per-link peak age $A^{\text{p}}(\pi_Q)/N$ achieved over a much larger time horizon.}\label{fig:main9}
\end{figure}

We now analyze performance of our proposed policies $\pi_Q$ and $\pi_A$ over the choice of parameters $V$ and $\beta$, respectively. Here, we set $K = 5$ and the number of `bad' channels also to be $n_{\text{bad}} = 5$. For the virtual-queue based policy $\pi_Q$, we observe that the parameter $V$ has nearly no effect on convergence time of the algorithm. To illustrate this, in Figure~\ref{fig:main9}, we plot per-link peak age $A^{\text{p}}(\pi_Q)/N$ computed over the first $t$ time slots, for two different values of $V = 0.1$ and $V = 100$. We observe that the peak age measured over the first $t$ slots converged to the peak age $A^{\text{p}}(\pi_Q)$ at nearly the same time.

\begin{figure}
  \centering
  \includegraphics[width=0.85\linewidth]{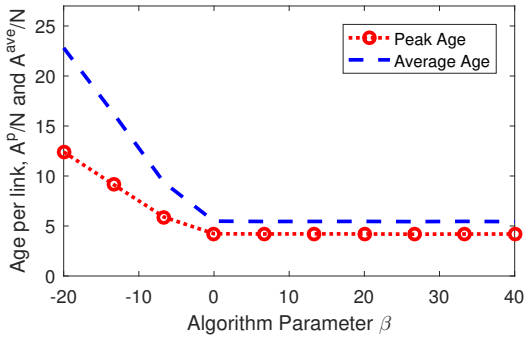}
  \caption{Per-link average and peak age, $A^{\text{ave}}(\pi_A)/N$ and $A^{\text{p}}(\pi_A)/N$, as a function of parameter $\beta$.}\label{fig:main8}
\end{figure}

For the age-based policy $\pi_A$, we again observe no difference in convergence time with respect to $\beta$. Theorem~\ref{thm:piA} guarantees bounds for any $\beta \in \mathbb{R}$. However, in Figure~\ref{fig:main8}, we observe that the peak and average age achieved by $\pi_A$ worsen as $\beta$ becomes negative. This is because $c_1(\beta)$ and $c_2(\beta)$ in Theorem~\ref{thm:piA} are large and negative when $\beta < 0$.

%Plot 2: Consider a $4\times4$ grid with $24$ links. We consider 1-hop constraints, i.e., matching constraints. Assume the channel process $\{ S_{e}(t) \}_t$ to be Markov chain, independent across links $e$, with $\pr{S_{e}(t) = 0 | S_{e}(t) = 0} = \pr{S_{e}(t) = 1 | S_{e}(t) = 1} = p$. Note that if $p = 1/2$ this is an i.i.d. process. $p \leq 1/2$ implies that the channel is more abrupt while if $p \geq 1/2$ then the channel is more stable. We plot the performance of proposed policies, and the age lower-bounds as a function of $p \in (0,1)$. Have to choose $V$ and $\beta$ for the two policies.

\section{Conclusion}
\label{sec:conclusion}
We considered the problem of age minimization for a wireless network under general interference constraints and time varying channels, when the channel state information is perfectly available to the scheduler to make scheduling decisions. We proposed a virtual-queue based policy and an age-based policy to minimize age. We proved that the virtual-queue based policy is nearly peak age optimal, up to a constant additive factor, and that the age-based policy is at most a factor $4$ away from age optimality.

Comparison with our previous work, which derived age optimal policies when the channel state information is not available to the scheduler, we demonstrate a $4$ fold improvement in age when the channel state information is available to the schedule in a particular network setting. This work, therefore, establishes the utility in obtaining or using the channel state information in scheduling to minimize age.

\bibliographystyle{ieeetr}
%\bibliography{../../../../PaperTrack/books-bib,../../../../PaperTrack/cvxalgo-bib,../../../../PaperTrack/aoi-bib,../../../../PaperTrack/uavnet-bib,../../../../PaperTrack/cps-bib,../../../../PaperTrack/neelesh-bib,../../../../PaperTrack/opt-scheduling-bib,../../../../PaperTrack/prob-bib}

\appendix

\subsection{Proof of Lemma~\ref{lem:time_conservation}}
\label{pf:lem:time_conservation}
Age evolution for link $e$ can be written as
\begin{equation}
A_{e}(t+1) = 1 + A_{e}(t) - U_{e}(t)S_{e}(t)A_{e}(t),
\end{equation}
for all $t$. As a result, we have
\begin{align}
A_{e}(t) - A_{e}(0) &= \sum_{\tau = 0}^{t-1} \left( A_{e}(\tau + 1) - A_{e}(\tau)\right), \nonumber \\
&= \sum_{\tau = 0}^{t-1} \left( 1 - U_{e}(\tau)S_{e}(\tau)A_{e}(\tau)\right), \nonumber \\
&= t -  \sum_{\tau = 0}^{t-1}U_{e}(\tau)S_{e}(\tau)A_{e}(\tau).
\end{align}
Since $A_{e}(0) = 0$, we have
\begin{equation}
\label{eq:t1}
\frac{1}{t}A_{e}(t) = 1 - \frac{1}{t} \sum_{\tau = 0}^{t-1}U_{e}(\tau)S_{e}(\tau)A_{e}(\tau).
\end{equation}
For $\pi \in \overline{\Pi}_1$, we have $\limsup_{t \rightarrow \infty} \frac{1}{t}\EX{A_{e}(t)} = 0$. Taking expected value on both sides of~\eqref{eq:t1}, and making $t \rightarrow \infty$, yields the result.

\subsection{Derivation of the Peak Age Minimization Problem}
\label{pf:der_peak_min_prob}
Using Lemma~\ref{lem:time_conservation}, we first show that for $\pi \in \overline{\Pi}_1$ is given by
\begin{equation}
\overline{A}^{\text{p}}_{e} = \frac{1}{\liminf_{t \rightarrow \infty} \EX{\frac{1}{t}\sum_{\tau = 0}^{t}\sum_{e \in E}U_{e}(\tau)S_{e}(\tau)}},
\end{equation}
for every $e \in E$. To see this, note that the peak age of link $e$ is given by
\begin{align}
\overline{A}^{\text{p}}_{e} &= \limsup_{t \rightarrow \infty} \frac{ \EX{\sum_{\tau = 0}^{t-1} U_{e}(\tau)S_{e}(\tau)A_{e}(\tau)} }{ \EX{\sum_{\tau = 0}^{t-1}U_{e}(\tau)S_{e}(\tau)} }, \nonumber \\
&= \limsup_{t \rightarrow \infty} \frac{  \EX{\frac{1}{t}\sum_{\tau = 0}^{t-1} U_{e}(\tau)S_{e}(\tau)A_{e}(\tau)} }{ \EX{\frac{1}{t}\sum_{\tau = 0}^{t-1}U_{e}(\tau)S_{e}(\tau)} }, \nonumber \\
&= \frac{ \limsup_{t\rightarrow \infty} \EX{\frac{1}{t}\sum_{\tau = 0}^{t-1} U_{e}(\tau)S_{e}(\tau)A_{e}(\tau)} }{ \liminf_{t \rightarrow \infty} \EX{\frac{1}{t}\sum_{\tau = 0}^{t-1}U_{e}(\tau)S_{e}(\tau)} }, \nonumber \\
&= \frac{1}{ \liminf_{t \rightarrow \infty} \EX{\frac{1}{t}\sum_{\tau = 0}^{t-1}U_{e}(\tau)S_{e}(\tau)} }, \label{eq:tani}
\end{align}
where the last equality follows from Lemma~\ref{lem:time_conservation}. Since $\overline{A}^{\text{p}}(\pi) = \sum_{e \in E}w_e \overline{A}^{\text{p}}_{e}(\pi)$, the peak age minimization problem $\min_{\pi \in \overline{\Pi}_1} \overline{A}^{\text{p}}(\pi)$ can now be written as
\begin{align}
\begin{aligned}
& \underset{\pi \in \overline{\Pi}_1}{\text{Minimize}} & & \sum_{e \in E}\frac{w_e}{\liminf_{t \rightarrow \infty} \frac{1}{t}\sum_{\tau = 0}^{t-1}U_{e}(\tau)S_{e}(\tau)}.
\end{aligned}
\end{align}
Using auxiliary variables $\alpha_e$, this can be written as~\eqref{eq:peak_min_problem}.

\subsection{Proof of Lemma~\ref{lem:ave_age_conservation}}
\label{pf:lem:ave_age_conservation}
We know that the age of link $e$ evolves as
\begin{equation}
A_{e}(t+1) = 1 + A_{e}(t) - U_{e}(t)S_{e}(t)A_{e}(t),
\end{equation}
for all $t$. Squaring this we obtain
\begin{multline}
\label{eq:s1}
A^{2}_{e}(t+1) = 1 + A^{2}_{e}(t) + U^{2}_{e}(t)S^{2}_{e}(t)A^{2}_{e}(t) + 2A_{e}(t) \\ - 2U_{e}(t)S_{e}(t)A^{2}_{e}(t) - 2U_{e}(t)S_{e}(t)A_{e}(t).
\end{multline}
Since $U_{e}(t)S_{e}(t) \in \{0, 1\}$, we have $U^{2}_{e}(t)S^{2}_{e}(t) = U_{e}(t)S_{e}(t)$. Substituting this in~\eqref{eq:s1} we get
\begin{multline}\label{eq:s2}
A^{2}_{e}(t+1) - A^{2}_{e}(t) = 1 + 2A_{e}(t) - U_{e}(t)S_{e}(t)A^{2}_{e}(t) \\ - 2U_{e}(t)S_{e}(t)A_{e}(t),
\end{multline}
for all $t$. Telescoping this over $t$ time slots we get
\begin{align}
A^{2}_{e}(t) - A^{2}_{e}(0) &= \sum_{\tau = 0}^{t-1} \left( A^{2}_{e}(\tau + 1) - A^{2}_{e}(\tau) \right), \nonumber \\
&= t + 2 \sum_{\tau = 0}^{t-1} A_{e}(\tau) -  \sum_{\tau = 0}^{t-1}U_{e}(\tau)S_{e}(\tau)A^{2}_{e}(\tau) \nonumber \\
&~~~~~~~~~- 2 \sum_{\tau = 0}^{t-1}U_{e}(\tau)S_{e}(\tau)A_{e}(\tau). \label{eq:s3}
\end{align}

Since the policy $\pi$ is in space $\overline{\Pi}_2$ we must have $\limsup_{t \rightarrow \infty} \frac{1}{t}\EX{A^{2}_{e}(t)} = 0$. Taking expectation in~\eqref{eq:s3}, using $\frac{1}{t}\EX{A^{2}_{e}(t)} \rightarrow 0$ and $A_{e}(0) = 0$, we get
\begin{align}
2\overline{A}^{\text{ave}}_{e} &= -1 + \limsup_{t \rightarrow \infty} \frac{1}{t}\EX{\sum_{\tau = 0}^{t-1}U_{e}(\tau)S_{e}(\tau)A^{2}_{e}(\tau)} \nonumber \\
&~~~~~+ 2\limsup_{t \rightarrow \infty} \frac{1}{t}\EX{\sum_{\tau=0}^{t-1}U_{e}(\tau)S_{e}(\tau)A_{e}(\tau)}, \nonumber \\
&= 1 + \limsup_{t \rightarrow \infty} \frac{1}{t}\EX{\sum_{\tau = 0}^{t-1}U_{e}(\tau)S_{e}(\tau)A^{2}_{e}(\tau)}, \label{eq:nnn333}
\end{align}
where the last equality follows from Lemma~\ref{lem:time_conservation}. This proves the Lemma for $\beta = 0$. From Lemma~\ref{lem:time_conservation}, we have that
\begin{equation}
0 = - 1 + \limsup_{t \rightarrow \infty} \frac{1}{t}\EX{\sum_{\tau = 0}^{t-1}U_{e}(\tau)S_{e}(\tau)A_{e}(\tau)}. \label{eq:nex}
\end{equation}
Adding $\beta$ times~\eqref{eq:nex} to~\eqref{eq:nnn333} we obtain the result, for any $\beta \in \mathbb{R}$.

\subsection{Proof of Theorem~\ref{thm:peak_age_lb}}
\label{pf:thm:peak_age_lb}
Let $\pi$ be a $\mathcal{S}$-only policy such that $\pi \in \overline{\Pi}_1$. Since, the channel process $\{ \mathbf{S}(t) \}_{t \geq 0}$ is i.i.d. across time $t$, the process $\{ U_{e}(t)S_{e}(t)\}_{t \geq 0}$ is also i.i.d. across $t$ for policy $\pi$, as $\mathbf{U}(t)$ is entirely determined by $\mathbf{S}(t)$. Therefore, we have $\alpha_e = \EX{U_{e}(t)S_{e}(t)}$ for all $t \geq 0$ and $e \in E$. Using Lemma~\ref{lem:time_conservation} and definition of peak age we have
\begin{align}
\overline{A}^{\text{p}}_{e} &= \frac{1}{\liminf_{t \rightarrow \infty} \frac{1}{t}\EX{\sum_{\tau=0}^{t-1}U_{e}(\tau)S_{e}(\tau)}}, \nonumber \\ &=\left\{ \begin{array}{ll}
  \frac{1}{\alpha_e} &~\text{if}~\alpha_e > 0 \\
  +\infty &~\text{if}~\alpha_e = 0
\end{array}\right.,
\end{align}
for all $e \in E$. Thus, the problem of peak age minimization over the space of all $\mathcal{S}$-only policies is equivalent to
\begin{align}
\begin{aligned}
&\underset{\bm{\alpha}}{\text{Minimize}}
& & \sum_{e \in E} \frac{w_{e}}{\alpha_e}, \\
& \text{subject to} && \bm{\alpha} \in \Lambda_{\mathcal{S}}\left( \bm{\gamma}\right).
\end{aligned}
\end{align}
The optimality of $\mathcal{S}$-only policies in solving~\eqref{eq:peak_min_problem} follows from Theorem~4.5 in~\cite{Neely_book}. This proves the result.

\subsection{Proof of Lemma~\ref{lem:peak_ave_bound}}
\label{pf:lem:peak_ave_bound}
Consider a policy $\pi \in \overline{\Pi}_2$, and a link $e$. From Cauchy-Schwartz inequality we have
\begin{multline}
\left( \EX{ \sum_{\tau=0}^{t-1}U_{e}(\tau)S_{e}(\tau)A_{e}(\tau) } \right)^{2} \leq \EX{ \sum_{\tau = 0}^{t-1} U_{e}(\tau)S_{e}(\tau) } \\
\times \EX{ \sum_{\tau = 0}^{t-1} U_{e}(\tau)S_{e}(\tau)A^{2}_{e}(\tau) }, \nonumber
\end{multline}
since $U^{2}_{e}(\tau)S^{2}_{e}(\tau) = U_{e}(\tau)S_{e}(\tau)$ as $U_{e}(\tau)S_{e}(\tau) \in \{0, 1\}$. Dividing both sides by $t^2$ we get
\begin{multline}
\frac{ \EX{ \left(\frac{1}{t}\sum_{\tau=0}^{t-1}U_{e}(\tau)S_{e}(\tau)A_{e}(\tau) } \right)^{2} }{ \EX{ \frac{1}{t}\sum_{\tau = 0}^{t-1} U_{e}(\tau)S_{e}(\tau) } } \\
\leq \EX {\frac{1}{t} \sum_{\tau = 0}^{t-1} U_{e}(\tau)S_{e}(\tau)A^{2}_{e}(\tau) }.
\end{multline}
Taking limsup on both sides and using Lemma~\ref{lem:time_conservation} and Lemma~\ref{lem:ave_age_conservation}, along with the definitions of $\overline{A}^{\text{p}}_{e}(\pi)$ and $\overline{A}^{\text{ave}}_{e}(\pi)$, we get
\begin{equation}
A^{\text{p}}_{e}(\pi) \leq 2A^{\text{ave}}_{e}(\pi) - 1.
\end{equation}
Summing over $e$ with weights $w_e$ we obtain the result.

In order to see that the inequality also holds at optimality, note that
\begin{align}
\overline{A}^{\text{p}\ast} = \inf_{\pi \in \overline{\Pi}_1} \overline{A}^{\text{p}}(\pi) &\leq   \inf_{\pi \in \overline{\Pi}_2} \overline{A}^{\text{p}}(\pi), \nonumber \\ &\leq \overline{A}^{\text{p}}(\pi) \leq 2\overline{A}^{\text{ave}}(\pi) - \sum_{e \in E}w_e, \label{eq:xxx}
\end{align}
for any $\pi \in \overline{\Pi}_2$, where the first inequality follows because $\overline{\Pi}_2 \subset \overline{\Pi}_1$. Taking infimum over $\pi \in \overline{\Pi}_2$ in~\eqref{eq:xxx} yields the result.

\subsection{Proof of Theorem~\ref{thm:piQ}}
\label{pf:thm:piQ}
\emph{Proof of Part~A}: Let $L(t) = \frac{1}{2}\sum_{e \in E}w_e Q^{2}_{e}(t)$ and $\Delta(t) = L(t+1) - L(t)$. Note that
\begin{align}
Q_{e}^{2}(t+1) &= \left[ \max\{ Q_{e}(t) + \alpha_{e}(t) - U_{e}(t)S_{e}(t), 1\} \right]^2, \nonumber \\
&\leq 1 + \left( Q_{e}(t) + \alpha_{e}(t) - U_{e}(t)S_{e}(t) \right)^2, \nonumber \\
&= 1 + \left( \alpha_{e}(t) - U_{e}(t)S_{e}(t) \right)^2 + Q^{2}_{e}(t) \nonumber \\
&~~~~~~~~~~~~~~~~~~+ 2Q_{e}(t)\left( \alpha_{e}(t) - U_{e}(t)S_{e}(t) \right), \nonumber \\
&\leq 1 + V + Q^{2}_{e}(t) + 2Q_{e}(t)\left( \alpha_{e}(t) - U_{e}(t)S_{e}(t) \right), \label{eq:r3}
\end{align}
where the last inequality follows from the fact that $\alpha_{e}(t) = \sqrt{\frac{V}{Q_{e}(t)}} \leq \sqrt{V}$ because $Q_{e}(t) \geq 1$ for all $t$.
Using~\eqref{eq:r3} we obtain
\begin{equation}
\label{eq:r4}
\Delta(t) \leq \frac{1+V}{2}\sum_{e \in E}w_e + \sum_{e \in E}w_e Q_{e}(t)\left( \alpha_{e}(t) - U_{e}(t)S_{e}(t) \right),
\end{equation}
for all $t$. We, therefore, have
\begin{multline}
Vg(\bm{\alpha}(t)) + \Delta(t) \leq V \sum_{e \in E}\frac{w_e}{\alpha_e(t)} + \frac{1+V}{2}\sum_{e \in E}w_e \\
+ \sum_{e \in E}w_eQ_{e}(t)\left[ \alpha_{e}(t) - U_{e}(t)S_{e}(t)\right]. \nonumber
\end{multline}
Substituting $\alpha_{e}(t) = \sqrt{V/Q_{e}(t)}$, which minimizes the right hand side, gives
\begin{multline}
Vg(\bm{\alpha}(t)) + \Delta(t) \leq \sum_{e \in E}2w_e\sqrt{VQ_{e}(t)} \\ + \frac{1+V}{2}\sum_{e \in E}w_e
- \sum_{e \in E}w_e U_{e}(t)S_{e}(t)Q_{e}(t). \label{eq:kk4}
\end{multline}
Policy $\pi_Q$ minimizes the right hand side of~\eqref{eq:kk4} as it activates set $m_{t}$ at $t$ which maximizes $\sum_{e \in m}w_e S_{e}(t)Q_{e}(t)$. Therefore, we can upper bound the right-hand side of~\eqref{eq:kk4} by the peak age optimal $\mathcal{S}$-only policy $\pi^{\ast}$:
\begin{multline}
Vg(\bm{\alpha}(t)) + \Delta(t) \leq \sum_{e \in E}2w_e\sqrt{VQ_{e}(t)} \\ + \frac{1+V}{2}\sum_{e \in E}w_e - \sum_{e \in E}w_e U^{\pi^{\ast}}_{e}(t)S_{e}(t)Q_{e}(t). \nonumber
\end{multline}
Since $\alpha^{\ast}_{e} = \EX{U^{\ast}_{e}(t)S_{e}(t)}$, taking conditional expectation in the above equation we get
\begin{multline}
\EX{Vg(\bm{\alpha}(t)) + \Delta(t)|\mathbf{Q}(t)} \leq \sum_{e \in E}2w_e\sqrt{VQ_{e}(t)} \\ + \frac{1+V}{2}\sum_{e \in E}w_e
- \sum_{e \in E}w_e \alpha^{\ast}_{e} Q_{e}(t), \label{eq:kk5}
\end{multline}
where $\bm{\alpha}^{\ast}$ is the solution to the peak age minimization problem in~\eqref{eq:age_opt}. This can be written as
\begin{multline}
\EX{Vg(\bm{\alpha}(t)) + \Delta(t)|\mathbf{Q}(t)} \leq V \overline{A}^{\text{p}\ast} + \frac{1+V}{2}\sum_{e \in E}w_e \\
- \sum_{e \in E}w_e \alpha^{\ast}_{e}\left[ \sqrt{Q_{e}(t)} - \frac{\sqrt{V}}{\alpha^{\ast}_{e}} \right]^2, \label{eq:kkk20}
\end{multline}
where $\overline{A}^{\text{p}\ast} = \sum_{e \in E}\frac{w_e}{\alpha^{\ast}_{e}}$ is the optimal value given in~\eqref{eq:age_opt}.

Now, ignoring the last term in~\eqref{eq:kkk20}, taking expected value, and summing both sides of~\eqref{eq:kkk20} over the first $t$ time slots we obtain
\begin{multline}
\EX{V \sum_{\tau=0}^{t-1} g(\bm{\alpha}(t))} + \EX{L(t) - L(0)} \\
\leq t\left[ V \overline{A}^{\text{p}\ast} + \frac{1+V}{2}\sum_{e \in E}w_e\right]. \nonumber
\end{multline}
Since $L(t) \geq 0$, we have
\begin{align}
\EX{V \sum_{\tau=0}^{t-1} g(\bm{\alpha}(t))} &\leq \EX{V \sum_{\tau=0}^{t-1} g(\bm{\alpha}(t))} + \EX{L(t)}, \nonumber \\
&\leq t\left[V \overline{A}^{\text{p}\ast} + \frac{1+V}{2}\sum_{e \in E}w_e\right] + \EX{L(0)}. \nonumber
\end{align}
Diving by $t$ and taking the limit we get
\begin{equation}
\limsup_{t \rightarrow \infty} \frac{1}{t}\EX{\sum_{\tau = 0}^{t-1}g(\bm{\alpha}(t))} \leq \overline{A}^{\text{p}\ast} + \frac{1}{2}\sum_{e \in E}w_e +  \frac{1}{2V}\sum_{e \in E}w_e. \label{eq:kk6}
\end{equation}
Since $g$ is convex, we have $g(\overline{\bm{\alpha}}(t)) \leq \frac{1}{t}\sum_{\tau = 0}^{t-1}g(\bm{\alpha}(t))$ from Jensen's inequality~\cite{boyd}. Substituting this in~\eqref{eq:kk6} yields the result.

\emph{Proof of Part~B}: Since $V g(\bm{\alpha}(t)) \geq 0$, from~\eqref{eq:kkk20} we obtain
\begin{equation}
\EX{\Delta(t)} \leq V\left[ \overline{A}^{\text{p}\ast} + \frac{1}{2}\sum_{e \in E}w_e \right] + \frac{1}{2}\sum_{e \in E}w_e.
\end{equation}
Summing this over $t$ time slots we get
\begin{equation}
\frac{1}{t}\EX{L(t)} \leq \frac{1}{t}\EX{L(0)} + V\left[ \overline{A}^{\text{p}\ast} + \frac{1}{2}\sum_{e \in E}w_e \right] + \frac{1}{2}\sum_{e \in E}w_e.
\end{equation}
This implies,
\begin{equation}
\label{eq:nd1}
\limsup_{t \rightarrow \infty} \frac{1}{t}\EX{L(t)} \leq B,
\end{equation}
where $B = V\left[ \overline{A}^{\text{p}\ast} + \frac{1}{2}\sum_{e \in E}w_e \right] + \frac{1}{2}\sum_{e \in E}w_e$. Now, since $L(t) = \frac{1}{2}\sum_{e \in E}w_e Q^{2}_{e}(t)$,~\eqref{eq:nd1} implies
\begin{equation}
\limsup_{t \rightarrow \infty} \frac{1}{t}\EX{Q^{2}_{e}(t)} \leq B,
\end{equation}
and as a consequence  $\limsup_{t \rightarrow \infty} \frac{1}{\sqrt{t}}\EX{Q_{e}(t)} \leq B$, for all $e \in E$, since $\EX{Q_{e}(t)}^2 \leq \EX{Q^2_{e}(t)}$. This implies
\begin{equation}
\limsup_{t \rightarrow \infty} \frac{1}{t}\EX{Q_{e}(t)} = 0,
\end{equation}
for all $e \in E$.

\emph{Proof of Part~C}: The queue evolution equation implies
\begin{equation}
Q_{e}(\tau+1) \geq Q_{e}(\tau) + \alpha_{e}(\tau) - U_{e}(\tau)S_{e}(\tau),
\end{equation}
for any $\tau \geq 0$. Summing this over $t$ times slots yields
\begin{equation}
\label{eq:w1}
\overline{\alpha}_{e}(t) + \frac{1}{t}Q_{e}(0) \leq \frac{1}{t}\sum_{\tau=0}^{t-1}U_{e}(\tau)S_{e}(\tau) + \frac{1}{t}Q_{e}(t),
\end{equation}
for all $t \geq 0$. Since $Q_{e}(t)$ is mean rate stable, taking expected value of~\eqref{eq:w1} and liminf as $t \rightarrow \infty$ we obtain
\begin{equation}
\liminf_{t \rightarrow \infty} \EX{\overline{\alpha}_{e}(t)} \leq \liminf_{t \rightarrow \infty} \frac{1}{t}\EX{\sum_{\tau=0}^{t-1}U_{e}(\tau)S_{e}(\tau)}.
\label{eq:w2}
\end{equation}

Since $g$ is a continuous, decreasing function in each $\alpha_e$ we have
\begin{align}
\overline{A}^{\text{p}}(\pi_Q) &= \sum_{e \in E} \frac{w_e}{\liminf_{t \rightarrow \infty} \EX{\frac{1}{t}\sum_{\tau = 0}^{t-1}U_{e}(t)S_{e}(t)}}, \nonumber \\
&\leq \sum_{e \in E} \frac{w_e}{\liminf_{t \rightarrow \infty} \EX{\overline{\alpha}_{e}(t)}}, \nonumber \\
&= \limsup_{t \rightarrow \infty} \sum_{e \in E} \frac{w_e}{\EX{\overline{\alpha}_{e}(t)}}, \nonumber \\
&\leq  \limsup_{t \rightarrow \infty} \EX{\sum_{e \in E} \frac{w_e}{\overline{\alpha}_{e}(t)} } = \limsup_{t \rightarrow \infty} \EX{g\left( \overline{\bm{\alpha}}(t)\right) },\label{eq:kk2}
\end{align}
where the first equality follows from Lemma~\ref{lem:time_conservation} and~\eqref{eq:peak_def}, the second inequality follows from~\eqref{eq:w2}, while the last inequality follows from Jensen's inequality~\cite{Durrett} and definition of $g(\bm{\alpha})$.

\subsection{Proof of Theorem~\ref{thm:piA}}
\label{pf:thm:piA}
Define $L(t) = \frac{1}{2}\sum_{e \in E} w_e A_{e}^{2}(t)$, $\Delta(t) = L(t+1) - L(t)$, and
\begin{multline}
\label{eq:q1}
f(t) =  \left( 1 - \beta\frac{(1-V)}{2}\right)\sum_{e \in E} w_e U_{e}(t)S_{e}(t)A_{e}(t) \\
+ \frac{V}{2}\sum_{e \in E} w_e U_{e}(t) S_{e}(t) A_{e}^{2}(t),
\end{multline}
for $0 < V < 1$, $\beta \in \mathbb{R}$, and all $t \geq 0$.
Using age evolution equation $A_{e}(t+1) = 1 + A_{e}(t) - U_{e}(t)S_{e}(t)A_{e}(t)$,
we obtain
\begin{multline}
\label{eq:q2}
\Delta(t) = \frac{1}{2}\sum_{e \in E}w_e + \sum_{e \in E}w_e A_{e}(t) \\
-\sum_{e \in E}w_e U_{e}(t)S_{e}(t)A_{e}(t) - \frac{1}{2}\sum_{e \in E}w_e U_{e}(t)S_{e}(t)A^{2}_{e}(t).
\end{multline}
Summing~\eqref{eq:q1} and~\eqref{eq:q2} we get
\begin{multline}
f(t) + \Delta(t) = \frac{1}{2}\sum_{e \in E}w_e + \sum_{e \in E} w_e A_{e}(t) \\ - \frac{(1-V)}{2}\sum_{e \in E}w_e U_{e}(t) S_{e}(t) \left[ A_{e}^{2}(t) + \beta A_{e}(t) \right].
\label{eq:nnn}
\end{multline}
Policy $\pi_A$ chooses $\mathbf{U}(t)$ that maximizes
\begin{equation}
\sum_{e \in E}w_e U_{e}(t) S_{e}(t) \left[ A_{e}^{2}(t) + \beta A_{e}(t) \right],
\end{equation}
and thus, it minimizes the right-hand side in~\eqref{eq:nnn}. Therefore, for any other policy $\pi$ we must have
\begin{multline}
f(t) + \Delta(t) \leq \frac{1}{2}\sum_{e \in E}w_e + \sum_{e \in E} w_e A_{e}(t) \\ - \frac{(1-V)}{2}\sum_{e \in E}w_e U^{\pi}_{e}(t) S_{e}(t) \left[ A_{e}^{2}(t) + \beta A_{e}(t)\right],
\end{multline}
where $\mathbf{U}^{\pi}(t)$ denotes the action of policy $\pi$ at time $t$. Substituting $\pi = \pi^{\ast}$, which is the peak age optimal $\mathcal{S}$-only policy that solves~\eqref{eq:age_opt}, gives the bound
\begin{multline}
\EX{f(t) + \Delta(t) \big| \mathbf{A}(t)} \leq \frac{1}{2}\sum_{e \in E}w_e + \sum_{e \in E} w_e A_{e}(t) \\
- \frac{(1-V)}{2}\sum_{e \in E}w_e \alpha^{\ast}_e\left[A_{e}^{2}(t) + \beta A_{e}(t)\right], \label{eq:last_one}
\end{multline}
since $\alpha^{\ast}_{e} = \EX{U^{\pi^{\ast}}_{e}(t)S_{e}(t)}$ is given by the solution to~\eqref{eq:age_opt}, and $U^{\pi^{\ast}}_{e}(t), S_{e}(t)$ are independent of $A_{e}(t)$ as $\pi^{\ast}$ is an $\mathcal{S}$-only policy. This can be re-written as
\begin{multline}
\EX{f(t) + \Delta(t) \big| \mathbf{A}(t)} \leq \frac{1}{2}\sum_{e \in E}w_e\\
+ \frac{1-V}{2}\sum_{e \in E} w_e \alpha^{\ast}_{e} \left[ \frac{\beta^2}{4} + \frac{(1-V)^{-2}}{\alpha^{\ast 2}_{e}} - \frac{1}{1-V}\frac{\beta}{\alpha^{\ast}_e }\right]\\
- \frac{(1-V)}{2}\!\sum_{e \in E}w_e \alpha^{\ast}_e \! \left[ A_{e}(t) + \frac{\beta}{2} - \frac{(1-V)^{-1}}{\alpha^{\ast}_{e} }\right]^2. \label{eq:star0}
\end{multline}
Ignoring the last term, since it is negative, and using the fact that $\alpha^{\ast}_{e} \leq 1$ we have
\begin{multline}
\EX{f(t) + \Delta(t)} \leq \frac{(1-V)^{-1}}{2}\sum_{e \in E}\frac{w_e}{\alpha^{\ast}_e} + \theta \sum_{e \in E}w_e,
\label{eq:star}
\end{multline}
where $\theta = \frac{1-\beta}{2} + (1-V)\frac{\beta^2}{4}$.
Summing this over $t$ time slots we obtain
\begin{multline}
\EX{\sum_{\tau = 0}^{t-1}f(\tau)} + \EX{L(t) - L(0)} \\
\leq t \left[ \frac{(1-V)^{-1}}{2}\sum_{e \in E}\frac{w_e}{\alpha^{\ast}_e} + \theta \sum_{e \in E}w_e \right].
\end{multline}
Since $L(t) \geq 0$ for all $t$, we have
\begin{align}
&\EX{\sum_{\tau = 0}^{t-1}f(\tau)} \leq  \EX{\sum_{\tau = 0}^{t-1}f(\tau)} + \EX{L(t)}, \nonumber  \\
&~~~~~~\leq t \left[ \frac{(1-V)^{-1}}{2}\sum_{e \in E}\frac{w_e}{\alpha^{\ast}_e} + \theta \sum_{e \in E}w_e \right] + \EX{L(0)}. \nonumber
\end{align}
Dividing this by $t$ and taking the limit we obtain
\begin{equation}
\limsup_{t \rightarrow  \infty} \frac{1}{t} \EX{\sum_{\tau = 0}^{t-1}f(\tau)} \leq \frac{(1-V)^{-1}}{2}\sum_{e \in E}\frac{w_e}{\alpha^{\ast}_e} + \theta \sum_{e \in E}w_e. \label{eq:star2}
\end{equation}
Note that $\overline{A}^{\text{p}\ast} = \sum_{e \in E}\frac{w_e}{\alpha^{\ast}_e}$, by Theorem~\ref{thm:peak_age_lb}. We also know from Lemma~\ref{lem:peak_ave_bound} that $\overline{A}^{\text{p}\ast} \leq 2\overline{A}^{\text{ave}\ast} - \sum_{e \in E} w_e$. Substituting this in~\eqref{eq:star2} we get
\begin{multline}
\limsup_{t \rightarrow \infty} \EX{\frac{1}{t}\sum_{\tau=0}^{t-1}f(\tau)} \leq \frac{1}{(1-V)}\overline{A}^{\text{ave}\ast} \\
+ \left( \theta - \frac{1}{2(1-V)}\right)\sum_{e \in E} w_e. \label{eq:star3}
\end{multline}

Assuming that $\EX{A_{e}^{2}(t)}$ is uniformly bounded for all $t$, we can make use of Lemma~\ref{lem:time_conservation} and~\ref{lem:ave_age_conservation} to compute $\limsup_{t \rightarrow \infty} \EX{\frac{1}{t}\sum_{\tau=0}^{t-1}f(\tau)}$. This gives us
\begin{multline}
\limsup_{t \rightarrow \infty} \EX{\frac{1}{t}\sum_{\tau=0}^{t-1}f(\tau)} = \sum_{e \in E} w_e + V\overline{A}^{\text{ave}}\left( \pi_A\right) \\
- \frac{\beta(1-V) + V}{2}\sum_{e \in E} w_e. \label{eq:nnt}
\end{multline}
Substituting this in~\eqref{eq:star3} we get
\begin{equation}
\overline{A}^{\text{ave}}\left( \pi_A\right) \leq \frac{1}{V(1-V)}\overline{A}^{\text{ave}\ast} - \kappa \sum_{e \in E} w_e,
\end{equation}
where $\kappa$ is given by
\begin{equation}
\kappa = \frac{1}{V} + \frac{1}{2V(1-V)} - \frac{\beta(1-V) + V}{2V} - \frac{\theta}{V}.
\end{equation}
Substituting $V = 1/2$ gives~\eqref{eq:thm:piA_1}.

In order to obtain~\eqref{eq:thm:piA_2}, notice that~\eqref{eq:star2} can be written as
\begin{equation}
\label{eq:nnt10}
\limsup_{t \rightarrow  \infty} \frac{1}{t} \EX{\sum_{\tau = 0}^{t-1}f(\tau)} \leq \frac{(1-V)^{-1}}{2}\overline{A}^{\text{p}\ast} + \theta \sum_{e \in E}w_e,
\end{equation}
since $\overline{A}^{\text{p}\ast} = \sum_{e \in E}\frac{w_e}{\alpha^{\ast}_{e}}$ due to Theorem~\ref{thm:peak_age_lb}. Now, using~\eqref{eq:nnt} and the fact that $\overline{A}^{\text{p}}(\pi_A) \leq 2\overline{A}^{\text{ave}}(\pi_A) - \sum_{e \in E}w_e$ from Lemma~\ref{lem:peak_ave_bound} we get
\begin{multline}
\label{eq:nnt11}
\limsup_{t \rightarrow  \infty} \frac{1}{t} \EX{\sum_{\tau = 0}^{t-1}f(\tau)} \geq \sum_{e \in E} w_e + \frac{V}{2}\overline{A}^{\text{p}}\left( \pi_A\right) \\
- \frac{\beta(1-V)}{2}\sum_{e \in E} w_e.
\end{multline}
Combining~\eqref{eq:nnt10} and~\eqref{eq:nnt11} in order to obtain a bound on $\overline{A}^{\text{p}}\left( \pi_A\right)$ as a function of $\overline{A}^{\text{p}\ast}$, and setting $V = 1/2$, we get the result in~\eqref{eq:thm:piA_2}.

It suffices to argue that the mean $\EX{A^{2}_{e}(t)}$ is uniformly bounded for all $t$. Define a Lyapunov function $\tilde{L}(t) = \frac{1}{2}\sum_{e \in E}w_e \left( A_{e}(t) + \beta/2 - 1\right)^2$, and the corresponding drift $\tilde{\Delta}(t) = \tilde{L}(t+1)  - \tilde{L}(t)$. Then using the same arguments as in~\eqref{eq:star0} we can obtain
\begin{equation}
\EX{\tilde{\Delta(t)}|\mathbf{A}(t)} \leq B_1 - \sum_{e \in E}B_{2,e}\left(A_{e}(t) + c_e \right)^{2},
\end{equation}
for constants $B_1$, $B_{2,e}$, and $c_e$. Foster-Lyapunov theorem~\cite[Chap.~6]{Hajek_rp} then implies that the process $\{ \mathbf{A}^{2}(t) \}_{t}$ is positive recurrent, and that $\EX{A^{2}_{e}(t)}$ is uniformly bounded.

\end{document}